\documentclass[amsmath,amssymb,twocolumn,superscriptaddress,aps,prl]{revtex4}

\usepackage[utf8]{inputenc}
\usepackage{amsthm}
\usepackage{adjustbox}
\usepackage{physics}
\usepackage{mathtools}
\usepackage{xcolor}
\usepackage{silence}
\usepackage{csquotes}
\usepackage{placeins}
\usepackage{graphicx}
\usepackage{soul}
\usepackage{bbold}
\usepackage{mathdots}
\usepackage{latexsym}
\usepackage{color}
\usepackage{color,soul}
\usepackage{nicefrac}
\usepackage{float}
\usepackage[T1]{fontenc}

\begin{document}

\title{Magnetocapacitance oscillations dominated by giant Rashba spin orbit interaction in InAs/GaSb quantum wells separated by AlSb barrier}

\author{A.S.L. Ribeiro}
\author{R. Schott}
\author{C. Reichl}
\author{W. Dietsche} 
\author{W. Wegscheider} 
\affiliation{Advanced Semiconductor Quantum Materials Group, Eidgenössische Technische Hochschule ETH, Zürich, Switzerland}
\affiliation{Quantum Center, ETH Zürich, Switzerland}
\date{\today}


\begin{abstract}
We observed magnetocapacitance oscillations in InAs/GaSb quantum wells separated by a $20$\,nm AlSb middle barrier. By realizing independent ohmic contacts for electrons in InAs and holes in the GaSb layer, 
we found an out-of-plane oscillatory response in capacitance representing the density of states of this system. We were able to tune the charge carrier densities by applying a DC bias voltage, identifying the formation of beating signatures for forward bias. The coexistence of two distinguishable two dimensional charge carrier systems of unequal densities was verified. The corresponding Landau phase diagram presents distinct features originating from the two observed densities. A giant Rashba coefficient ranging from $430-612$\,meV$\text{\AA}$ and large \textit{g}-factor value underlines the influence of spin orbit interaction.
\end{abstract}

\maketitle
\section{introduction}

The inverted band alignment present in InAs/GaSb quantum wells (QWs) provide a rich platform to investigate bulk and edge states properties \cite{PhysRevLett.115.036803,PhysRevLett.78.4613,du2017evidence,zakharova_PhysRevB.64.235332,edgestates_PhysRevB.97.045420,helicaledege_PhysRevLett.114.096802,Edgechannel_PhysRevB.87.235311,nichele2016edge,susanne_PhysRevB.96.075406}. Tunable via electric field, the properties of confined electrons and holes in the InAs/GaSb QWs have been studied in the past years \cite{lapushkin2004self,PhysRevB.94.045317,padilla2018confinement,PhysRevLett.115.036803,LL_PhysRevB.69.115319,sipahi_PhysRevB.104.195307}. Arising from the band inversion for corresponding QW layer thicknesses, the predicted quantum spin Hall effect insulators (QSHIs) \cite{QSHIproposalPhysRevLett.96.106802,Interplay_QSHI_PhysRevB.106.235421,QSHI_HgTe_InAs/GaSb,PhysRevLett.100.236601,Muraki_QSHI_10.1063_1.4967471,SHE_PhysRevLett.100.056602}, known as two dimensional topological insulators ($2$DTI), have been demonstrated \cite{Interplay_QSHI_PhysRevB.106.235421,QSHI_HgTe_InAs/GaSb,du2013observation}. The interlayer tunability is achieved by means of gate control, which offers the possibility of transitioning from a non trivial TI semimetal towards a conventional semiconductor \cite{tunable_gates_10.1063/1.118187,nguyen2015high,Irie_PhysRevMaterials.4.104201}. 

Given the inverted band alignment, the lowest conduction band (CB) energy level for electrons in InAs is spatially separated from and lies below the highest valence band (VB) hole states in GaSb. Adding an AlSb barrier in between InAs and GaSb layers, this unique arrangement offers the chance to investigate the tunability and coupling of $2$D electron-hole correlated systems, depending on the thickness of the AlSb layer separating them \cite{naveh_barrier,naveh1995band,ZHU1990595}. This scenario opens the opportunity to create a bound excitonic state with overlapping wave functions along the interface of InAs and GaSb with the AlSb barrier, enabling the potential Bose-Einstein condensation \cite{Eisenstein2004BoseEinsteinCO,du2017evidence,laikhtman_de1999band,hirayama_PhysRevB.67.195319}. Due to the insulating barrier, this device geometry with broken band gap can be associated with a plate capacitor, where each plate is charged with electrons and holes separately. In addition, the inserted middle AlSb barrier offers the possibility of tuning the coupling of both charge carrier types and, as a consequence tailoring the hybridization gap. 

\begin{figure}[htbp]
        \centering
    	\includegraphics[width = 0.95\columnwidth]{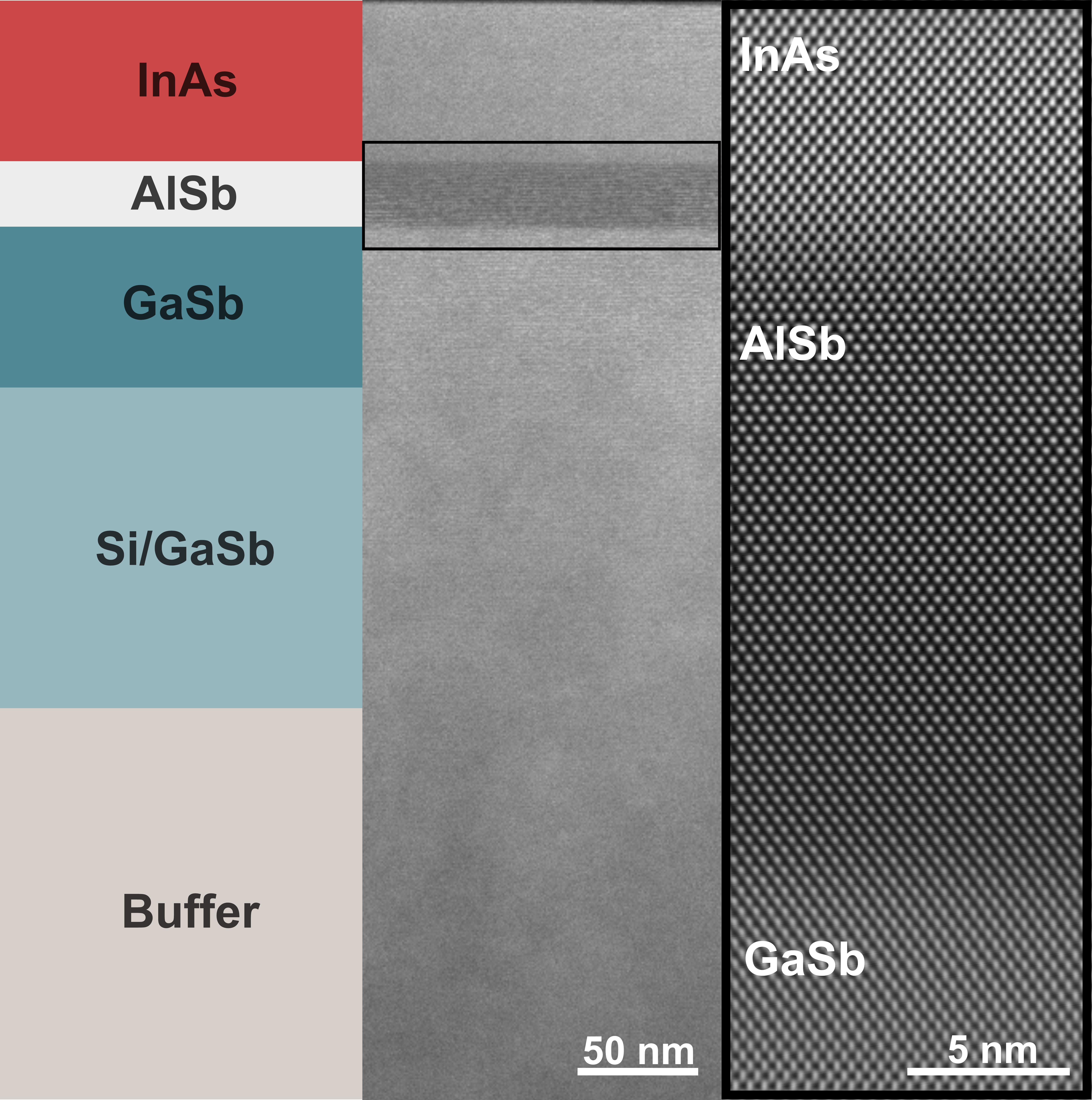}
	\caption{\textbf{Layer structure and STEM images.} Displayed from left to right: the layer structure of the grown InAs/AlSb/GaSb heterostructure, next to scanning transmission electron microscopy (STEM) images in different magnification scales (middle and right panel). The middle panel shows an overview image of the layer structure from the semiconductor surface down to the beginning of the buffer layer, according to the left most panel. The AlSb barrier surrounded by the InAs/GaSb quantum well interfaces, denoted with a black rectangle, is depicted with atomic resolution in the right most panel.}
	\label{fig: layer structure}
\end{figure}

In this work, we realize for the first time magnetocapacitance investigations in such a InAs/GaSb system separated by a $20$\,nm AlSb insulating barrier. In contrast to previous in-plane Shubnikov deHaas (SdH) transport, we introduce an out-of-plane measurement technique. By employing independent ohmic contacts to each channel, we obtain a capacitive response, revealing an unconventional Landau phase diagram with tunable densities modulated by a DC bias voltage. The spin orbit interaction (SOI) in InAs and GaSb layers influences the electrical properties of charge carriers such as giant Rashba spin splitting and enhanced \textit{g}-factor values. The presence of the hybridization between spin up and spin down states emerges as the dominant effect in this InAs/AlSb/GaSb system.

\begin{figure}[htbp]
        \centering
	   \includegraphics[width = 0.8\columnwidth]{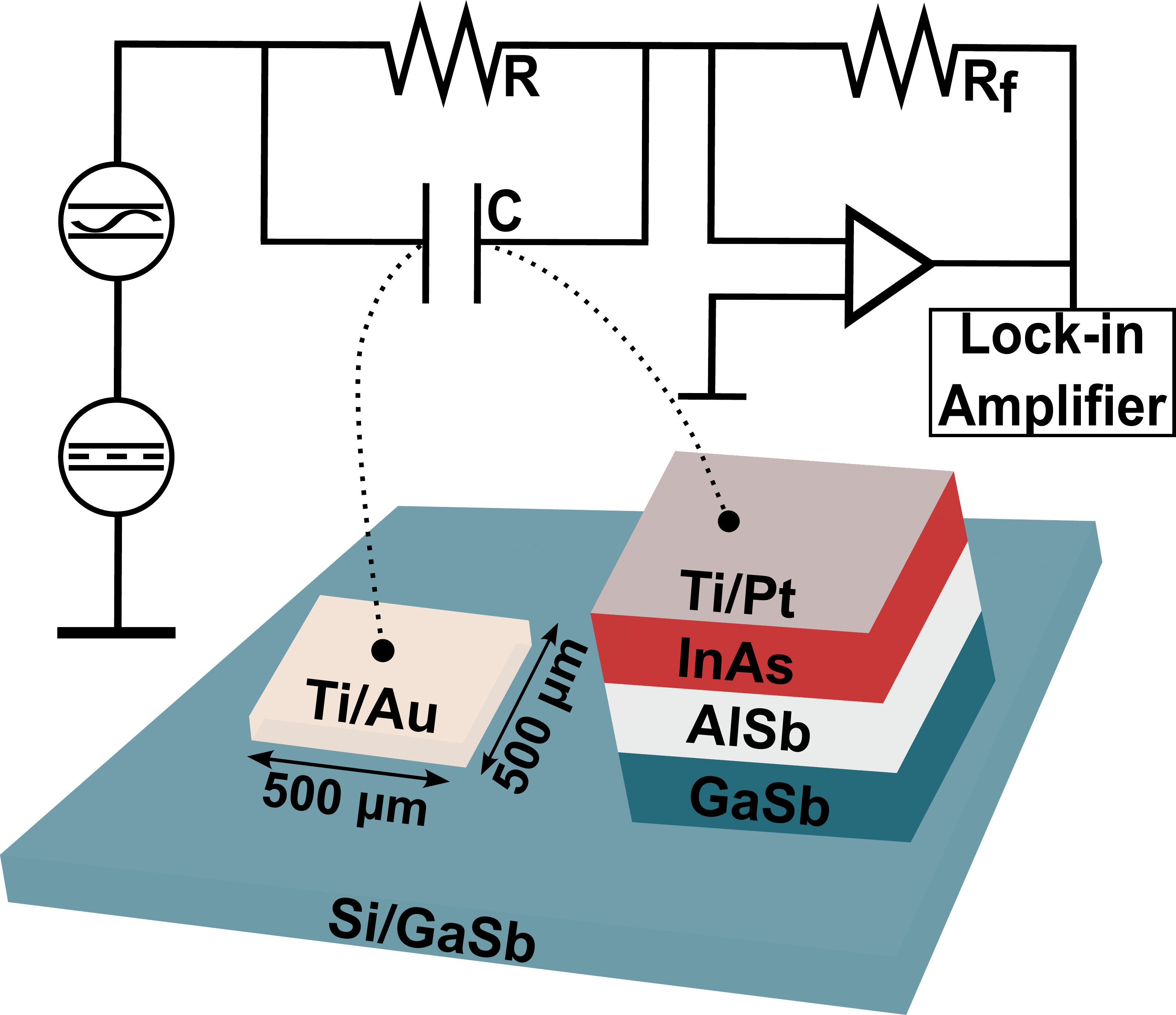}
	\caption{\textbf{Capacitance measurement setup with sample geometry}. The upper schematic displays the setup used for the magnetocapacitance experiment. The pillar geometry on the bottom represents the out-of-plane measurement technique, which allows to contact holes in the GaSb layer and electrons in the InAs layer.} 
	\label{fig: Ccircuit with sample}
\end{figure}

\section{Experimental approach}

We consider a symmetric heterostructure constituted of $50$\,nm of InAs and GaSb quantum wells, separated by a $20$\,nm AlSb barrier, grown by molecular beam epitaxy in the ($001$) crystal orientation, using an undoped GaSb substrate, as displayed on the left panel of Fig. \ref{fig: layer structure}. For further growth details, please refer to the Supplemental Material. 

A lamella film was prepared via the Focused Ion Beam (FIB) method. Scanning transmission electron microscopy (STEM) images taken at different magnifications are displayed in the middle and right panel of Fig. \ref{fig: layer structure}. In the center, we see an overview image covering the whole layer stack corresponding to the left most panel enclosing the InAs surface towards the beginning of the buffer layer. In the right most panel, we display a zoomed in image denoted by the black rectangle area in the middle panel of Fig. \ref{fig: layer structure}. We clearly see that the InAs/AlSb/GaSb interfaces are free of defects and that intermixing is negligible. This is important for the reliability of our heterostructure with respect to the insulation of the AlSb barrier.

Conventional optical lithography fabrication was performed to define the mesa pattern in a pillar geometry, facilitating the capacitance measurement, as illustrated in Fig. \ref{fig: Ccircuit with sample}. Individual contacts to InAs and GaSb quantum wells were realized using wet chemical etching with a $H_2O:C_6H_8O_7:H_3PO_4:H_2O_2$ solution, with an etching rate of $0.4$\,nm/s for the present $20$\,nm wide AlSb barrier. To provide an ohmic contact, we metalized Ti/Au: $50/150$\,nm on the GaSb - p type - layer. For the InAs - n type - surface, a thin layer of Ti/Pt: $15/5$\,nm was evaporated. A silver epoxy paste was used to connect Au wires to the n type region, while In soldering was used for the p type region. The I-V traces from p to n layer demonstrate a diode-like behavior ranging from $-0.1<V_{DC}<+0.1\,V$ (see Suppl. Mat.), exhibiting a central insulating low conductive region, of maximal resistance $\approx100$\,k$\Omega$.

Using standard Lock-in techniques coupled with an AC/DC voltage adder and an I-V converter connected directly to the Lock-in input signal, as depicted in Fig. \ref{fig: Ccircuit with sample}, we performed magnetocapacitance measurements with DC bias applied from p (GaSb) to n (InAs) layer as a function of perpendicular magnetic field, using a frequency of $1.43$\,kHz. This frequency, provided the best signal to noise ratio quality of the out-of-phase signal output required for the magnetocapacitance measurements.

\begin{figure}[htbp]
    \centering
        \includegraphics[scale=0.53]{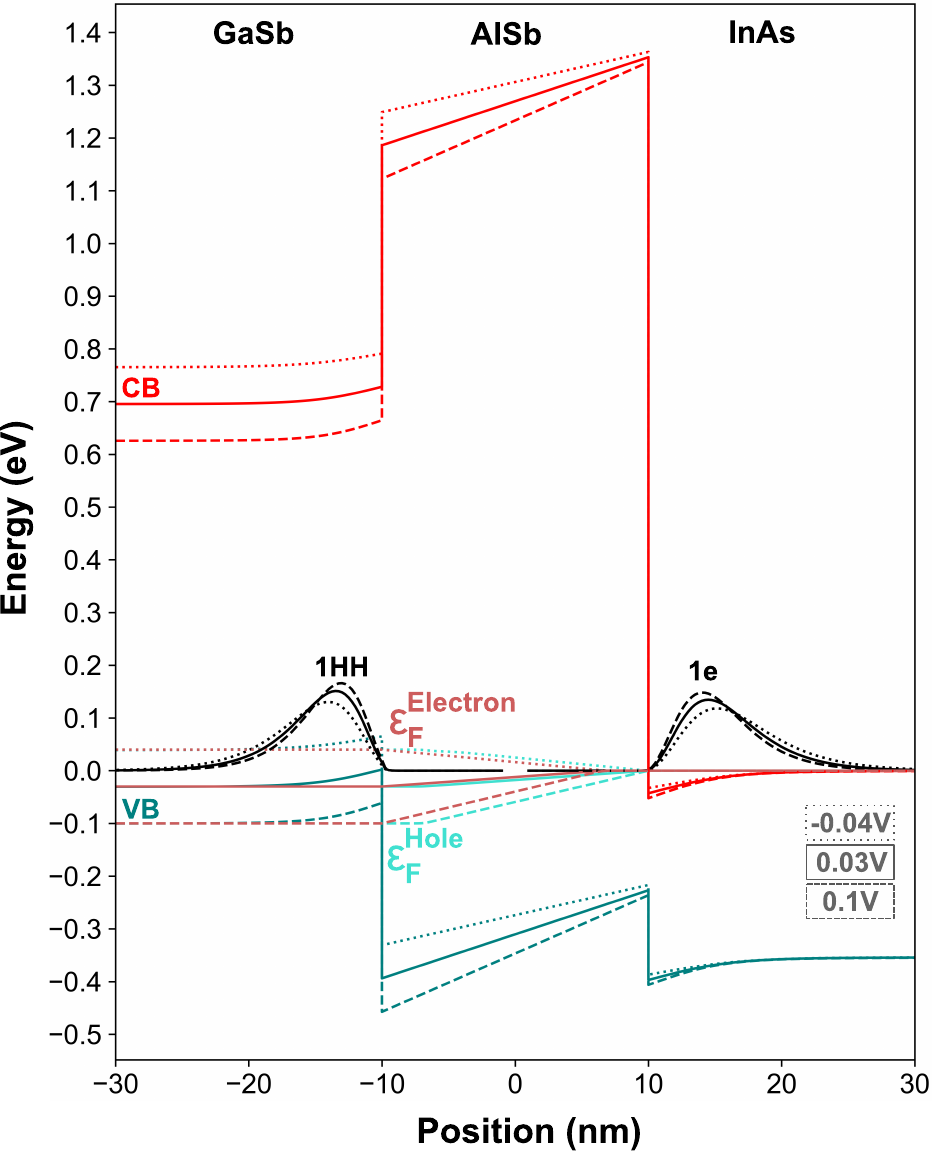}
    \caption{\textbf{Band structure simulation.} The band alignment for the grown InAs/AlSb/GaSb heterostructure for $-0.04$\,V (dotted lines), $0.03$\,V (solid lines) and $0.1$\,V (dashed lines) bias voltage applied from the p (GaSb) to the n (InAs) region are shown. The conduction band (CB) in red and the valence band (VB) in teal color for heavy holes are depicted, in addition to the quasi-Fermi energy levels \textbf{$\varepsilon_{F}$} for electrons (light brown) and holes (turquoise) for both bias polarities. The ground state probability density distribution of electrons in InAs ($1$e) and heavy holes in GaSb (1HH) are illustrated by black curves.}
    \label{fig:Band structure nextnano}
\end{figure}


\section{results}

We simulate the band structure alignment around the $20$\,nm AlSb barrier in between InAs and GaSb layers, by solving the $1$D Schrödinger and Poisson equations in the Nextnano platform \cite{nextnano4294186,nextnano_article}. Results for different applied bias voltage are depicted in Fig. \ref{fig:Band structure nextnano}. Due to the expected inverted band alignment, we see the confinement of electrons in the InAs and holes in the GaSb layers. The probability densities of electrons ($1$e) and heavy holes ($1$HH) in the ground state predict the formation of a $2$-Dimensional electrons gas ($2$DEG) and a $2$-Dimensional hole gas ($2$DHG), respectively, within the corresponding quantum wells.

In equilibrium regime, where no bias voltage is applied, the simulation certifies the existence of a built-in electric field. We note that for the non equilibrium regime, 
the confinement of carriers remains and the band alignment is modified as the Fermi energy level ($\varepsilon_{F}$) is modulated for forward and reverse bias. It enables the tuning of charge carrier densities in both channels simultaneously. This shifts the position of the electrochemical potential of the conduction and valence bands, respectively described by the quasi-Fermi energy for electrons and holes. This variation causes a split in both quasi-Fermi levels, as seen in the simulation (Fig. \ref{fig:Band structure nextnano}), under both forward and reverse DC bias polarity. The density probability of holes in the GaSb quantum well becomes non localized for a reverse bias larger than $-0.04$\,V, not shown in Fig. \ref{fig:Band structure nextnano} for clarity.

\begin{figure}[htbp]
    \centering
        \includegraphics[scale=0.8]{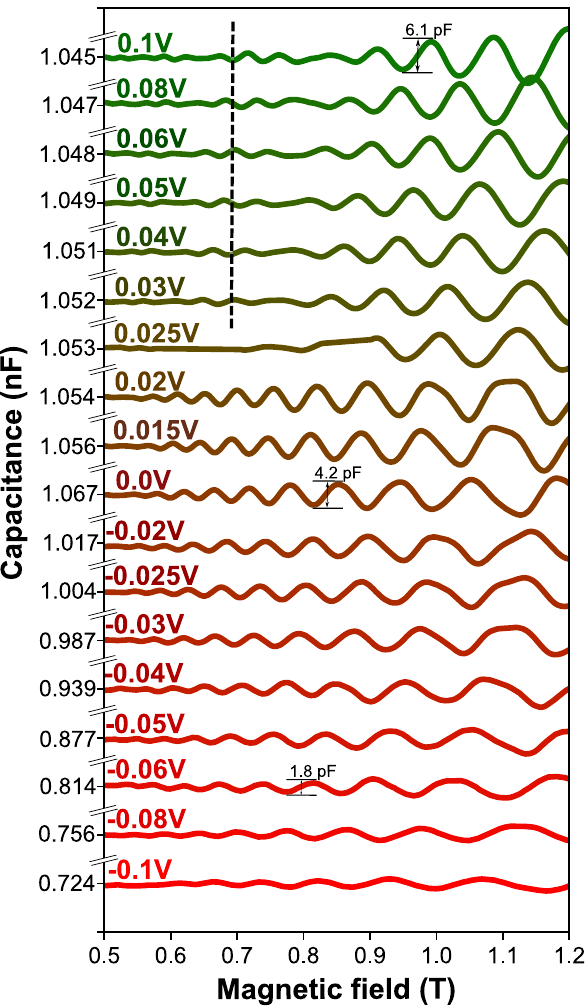}
    \caption{\textbf{Carrier density tunability.} Magnetocapacitance signatures for forward and reversed DC bias modulation measured in the low magnetic field range (without Zeeman spin splitting). The black dashed line indicates the occurrence of beating signatures for forward bias. The amplitudes in capacitance values for a few curves are depicted for comparison.}
    \label{fig:C vs B all dc bias}
\end{figure}

In Fig. \ref{fig:C vs B all dc bias}, we present magnetocapacitance oscillations performed in a dilution refrigerator at a base temperature of $25$\,mK. We clearly identify the similarity to an in-plane longitudinal resistivity behavior, i.e., Shubnikov deHaas (SdH) oscillations, with the presence of Zeeman spin splitting (roughly at $1.1$\,T) already for low magnetic field values. Here, we note that the oscillations appear for an out-of-plane scheme, due to the measurement design itself. The emergence of a beating pattern for forward bias voltage larger than $0.02$\,V is well pronounced, pointed out with a dashed black line, for clarity. Apart from the usual periodicity in $1$/B, this is a typical signature of two different densities leading to oscillations with different frequencies. The capacitive signal amplitude diminishes for reversed bias. This can be explained with an increase of fluctuating potentials and consequently disorder effects, contributing to enhanced scattering for the lower carrier densities. 

Magnetocapacitance oscillations are a direct measurement of the quantum capacitance. The latter originates from the modified density of states (DOS), resulting from the quantization of Landau levels for a perpendicular magnetic field. According to the equation for the effective Capacitance $C_{eff}$:

\begin{equation}
    \frac{1}{C_{eff}}=\frac{1}{C_{g}} + \frac{1}{C_{q}}
\end{equation}
the contribution of the geometrical and quantum capacitance, $C_{g}$ and $C_{q}$, respectively, has to be considered. 

In the absence of a magnetic field, the first term dominates as $C_{q}$ reaches huge values of the order of $5.74 \times 10^{35}$\,F/A, according to the definition DOS$_{2D}=\mu/\pi\hbar^{2}$, where $\mu$ and $\hbar$ denotes the reduced effective mass and the reduced Planck constant, respectively. Considering the reduced effective mass contribution of electrons in InAs (m$^{*}=0.023m_{o}$) and holes in GaSb (m$^{*}=0.4m_{o}$), the finite DOS in the $2$D gas corresponds to the length scale given by the effective Bohr radius $a_{B}^{*}$, which increases the effective separation of the capacitor plates by $a_{B}^{*}/4\approx9.03$\,nm \cite{ihn2009semiconductor}. This value is in good agreement with the simulation presented in Fig. \ref{fig:Band structure nextnano}, where we see that both wave function maxima ($1$HH and $1$e) are separated from the AlSb interface by practically the same length. The finite values depicted for each bias curve in Fig. \ref{fig:C vs B all dc bias}, are the constant zero field DOS where the geometrical capacitance dominates. Once we apply the magnetic field, quantization of Landau levels appears and we observe the oscillatory behavior. 

\begin{figure}[tbp]
    \centering
        \includegraphics[scale=0.58]{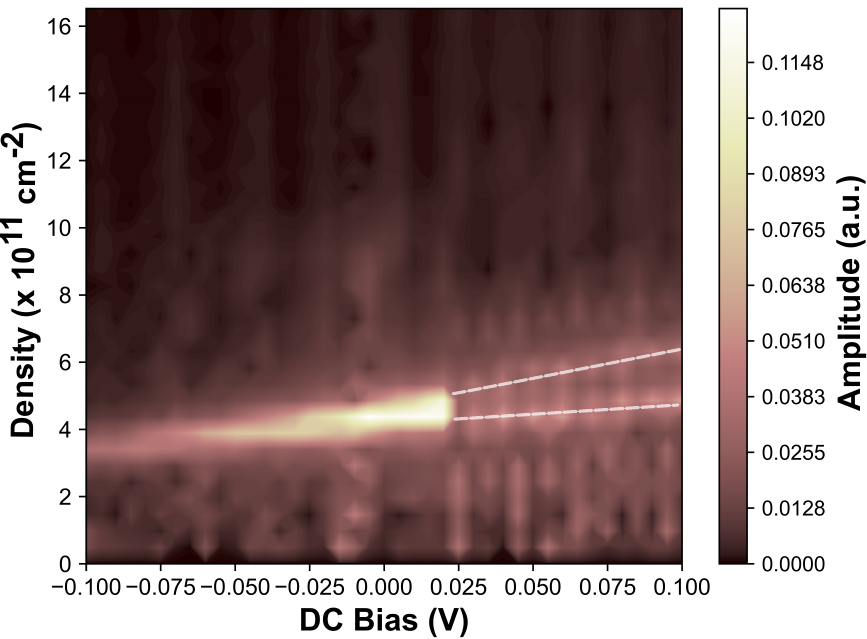}
    \caption{\textbf{Magnetocapacitance density evolution without Zeeman spin splitting signatures.} Power spectra calculated by the FFT of the magnetocapacitance curves at various DC bias voltage applied (as in Fig. \ref{fig:C vs B all dc bias}), determined for magnetic field up to one Tesla, addressed as case (I). The abrupt transition around $0.02$\,V is correlated to the imminence of beating signatures for forward bias, presented previously. The white dashed lines serves as an eye guide.}
    \label{fig: FFT fan no Zeeman Split}
\end{figure}

In order to extract charge carrier densities for each DC bias in the magnetocapacitance traces shown in Fig. \ref{fig:C vs B all dc bias}, a Fast Fourier transform (FFT) analysis was performed. We analyse the power spectra considering two cases according to different magnetic field ranges. In Fig. \ref{fig: FFT fan no Zeeman Split}, we show magnetocapacitance oscillations disregarding the Zeeman spin splitting (ZSS) until approximately $1$\,T and refer to case (I). In Fig. \ref{fig: FFT fan full range}, we display the whole range until $7$\,T, regarding as case (II). For case (I), the FFT demonstrates the existence of a single density channel for reversed and forward bias voltages up to $0.025$\,V. For higher forward bias, the FFT analysis reveals two non-parallel lines (marked with dashed lines in Fig. \ref{fig: FFT fan no Zeeman Split}), which correspond to beating signatures depicted already in Fig. \ref{fig:C vs B all dc bias}. 

\begin{figure}[tbp]
    \centering
        \includegraphics[scale=0.6]{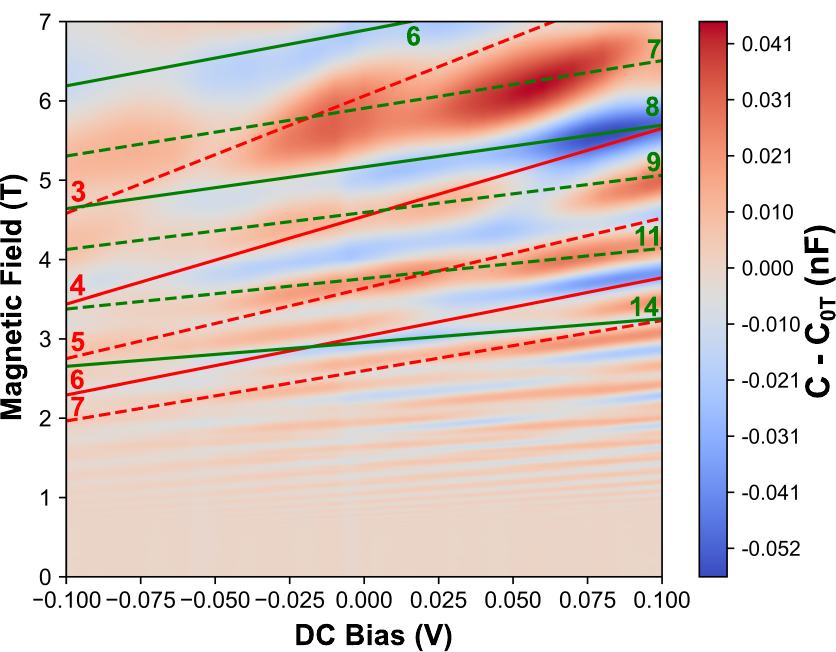}
    \caption{\textbf{Landau phase diagram with expected filling factors}. The magnetocapacitance oscillations as a function of the applied DC bias for the complete magnetic field range are shown. The colorbar indicates the difference between the measured capacitance values and the offset value at zero magnetic field for the respective DC bias curve. We display also the calculated filling factor positions from a partial range of the Landau phase diagram for clarity. The lines represent the predicted occupation of the first (red) and second (green) channel, where full (dashed) lines correspond to even (odd) filling factors.}
    \label{fig: L fan with ff expectation}
\end{figure}

Before displaying the power spectra for case (II), we show in Fig. \ref{fig: L fan with ff expectation}, the complete magnetocapacitance measurements for various applied DC bias voltages. We plot the expected filling factors ($\nu$) values as well, extracted from the power spectra analysis shown in Fig. \ref{fig: FFT fan full range} and remark that the ones for the lower density, named here as n$_{1}$, are displayed in red lines. The dashed or solid lines distinguish the parity of the filling factors. Analogously, the $\nu$ values correlated with the higher density, called n$_{2}$, are illustrated in green with the respective parity distinction likewise. The texture of our phase diagram is in good agreement with expected filling factor positions, showing the solid overlap with our FFT results. The lines in Fig. \ref{fig: L fan with ff expectation} (for both parities) correlated to one density, follow the areas with either lower or higher variance values in capacitance, according to the respective figure colorbar. 

Along this direction, we note the crossing between (anti) symmetric parity from n$_{1}$ and n$_{2}$. This feature is observed in two configurations: whenever an even to odd (and vice-versa) filling factors lines overlap and even to even (odd to odd) as well, as seen in Fig. \ref{fig: L fan with ff expectation}, indicating the intersection between electron and hole-like state \cite{phase_slips_Matija_PhysRevB.99.201402}. 

The effect of the applied DC bias voltage implies that the electric field between both capacitor plates tunes both carrier densities simultaneously, as seen in the power spectra depicted in Fig. \ref{fig: FFT fan full range} and predicted in the simulation displayed previously in Fig. \ref{fig:Band structure nextnano}. We see that we have two distinguishable curves correlated to the charge carrier density from each capacitor plate which are not multiple of each other. The linear correlation of density as a function of DC bias is illustrated in the same plot (Fig. \ref{fig: FFT fan full range}), with densities extracted from the power spectra associated with each DC bias applied. 

An important factor to consider from the linear fit in Fig. \ref{fig: FFT fan full range}, is that the slope directly represents, the capacitance per unit area A. By using the geometrical capacitance formula -- $\varepsilon\varepsilon_{o}A/d$,  where $\varepsilon$ and $\varepsilon_{o}$ denote the electrical permittivity of AlSb and vacuum respectively -- we were able to extract the distance d between the capacitor plates amounting to $55\pm 3$ nm. According to our simulation, we note that the wave functions maxima are separated by approximately $30$\,nm, which is in good agreement with the expected contribution considering the $20$\,nm AlSb barrier and the effective separation of the capacitor plates calculated previously. The remaining difference towards the experimentally amounted distance, we associate with the parasitic capacitance from our measurement setup.

\section{Rashba Spin Orbit Interaction}

\begin{figure}[tbp]
    \centering
        \includegraphics[scale=0.6]{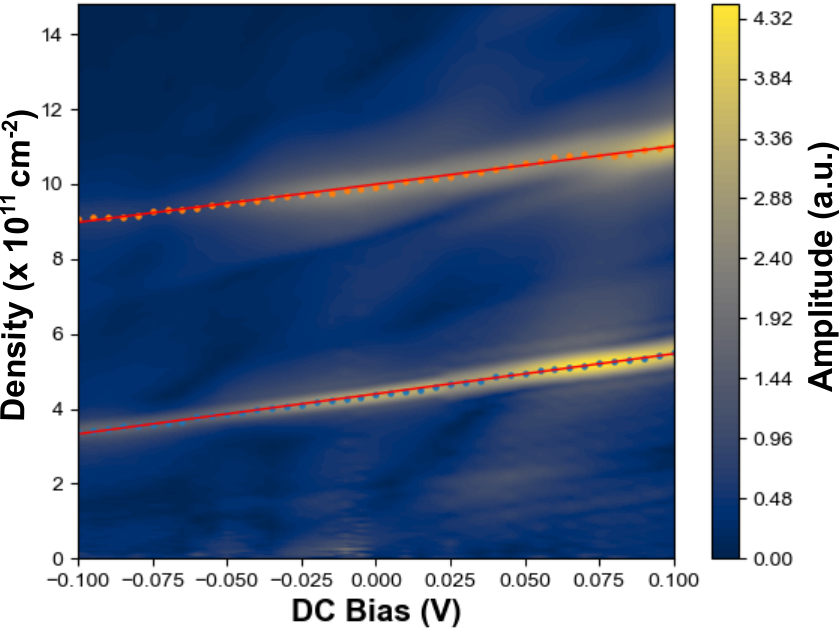}
    \caption{\textbf{Density evolution with Zeeman spin splitting.} Fast Fourier Transform of the complete magnetocapacitance range as a function of DC bias. The clear definition of both curves reveals the coexistence of two $2$D gases in the system. From the FFT analysis for all DC bias voltage applied, the respective densities were determined and plotted as a function of DC bias. Red lines indicate the linear fit of both curves.}
    \label{fig: FFT fan full range}
\end{figure}

Here, we describe the procedure to determine the Rashba coefficient $\alpha$, correlated to spin splitting due to structural inversion asymmetry (SIA). By considering the DC bias range where the beating signatures appear and the corresponding FFT spectra, we extracted the difference in charge carrier densities $\Delta{n}$ from Fig. \ref{fig: FFT fan full range}. According to Engels et al. \cite{Rashba_PhysRevB.55.R1958}, we obtained the Rashba coefficient $\alpha$ by the following equation 

\begin{equation}
    \alpha=\frac{\hbar^2\Delta{n}}{m^{*}}\sqrt{\frac{\pi}{2(n_{tot}-\Delta{n})}}
\end{equation}
where $m^{*}$, $\hbar$ and $n_{tot}$ represent the effective mass, reduced Planck constant and total density, respectively. As a result, we report a giant Rashba coefficient $\alpha$ varying from $\alpha = (430-612)$\,meV$\text{\AA}$. 

We note that Rashba spin orbit coupling should exhibit anti symmetric features as a function of in-plane momentum k$_\parallel$ for InAs/AlSb/GaSb heterostructures, as predicted by Li et al. \cite{spinstates_PhysRevB.80.035303}. The electron-hole like hybridization regime and the coupling mechanism differs from the InAs/GaSb quantum well as a consequence. As the Fermi energy $\varepsilon_{F}$ is located within the band gap, shown in our simulations (Fig. \ref{fig:Band structure nextnano}), the unexpected giant Rashba value we present here, serves as a good hint of the influence of spin-related properties into this InAs/AlSb/GaSb system.

Li et al. \cite{spinstates_PhysRevB.80.035303} show that the Rashba spin splitting (RSS) exhibits non linear and oscillatory behavior as a function of in-plane momemtum k$_\parallel$. It exists already in the absence of external electric field and arises from the asymmetry of conduction and valence subbands. The non linearity emerges from interface contributions and the oscillatory behavior is an outcome of the strong intermixing response between conduction and heavy-hole subbands. These features can be enhanced with external electric field. The sharp drop in RSS value for k$_\parallel=0$ shown in \cite{spinstates_PhysRevB.80.035303}, indicates the anticrossing point between conduction and heavy-hole subbands accompanied with a sign change. This sign change is related to the crossing of two spin branches, the spin up and down one, resulting in an anticrossing of spin states \cite{RSS_10.1063/1.2909544}. The spin orientation without the AlSb barrier deviate strongly along the Fermi circle due to the hybridization between both spin states. With the middle barrier, the hybridization is reduced and spin orientation diverge very little along the Fermi circle. This means that inserting the barrier, one can tune the spin orientations near the Fermi level by changing the thickness of the AlSb barrier. According to the authors \cite{spinstates_PhysRevB.80.035303,RSS_10.1063/1.2909544}, as the thickness of AlSb barrier is increased, the anticrossing between both subbands decreases and the hybridization gap narrows due to the suppression of charge carrier tunneling. This reduces the spin splitting of each subband as the structural inversion asymmetry decreases. For our case, this response differs in the sense that we have traces of tunneling happening for both bias polarities at approximately $\pm0.1$\,V (see Suppl. Mat.). In addition to that, the SIA seems to remain with and without bias applied, which keeps the spin splitting effect into play, in good agreement with the magnetocapacitance oscillations observed in Fig. \ref{fig:C vs B all dc bias}. These concepts can be translated to our Landau phase diagram (Fig. \ref{fig: L fan with ff expectation}), as we depict magnetocapacitance as a function of DC bias. Considering the giant Rashba coefficient we present and due to the flip in magnitude of capacitance (as shown by the colorbar), we assume that the distinct texture is associated with spin properties. Whenever there is a crossing between filling factors related to the densities n$_{1}$ and n$_{2}$, either for the same parity or not, there is an (anti) crossing between the two spin branches associated to the carriers in each capacitor plate.

A further physical quantity that is of great interest is the \textit{g}-factor. By recording the magnetic field where the first magnetocapacitance oscillation, named as $B_{c}$ and first spin splitting ($B_{ss}$) is observed in Fig. $4$, the \textit{g}-factor is calculated as \textit{g}$=2(B_{c}m_{o}/B_{ss}m^{*})$, with $m_{o}$ as the bare electron mass. This yield a variation of \textit{g} between $(23-45)$ as a function of DC bias (See Suppl. Mat.). This range is due to the uncertainty of $B_{c}$ values, which is greater than previously documented for an InAs quantum well using capacitance techniques \cite{gfactor_capacitance_Irie_2019}.


\section{Conclusion and Outlook}

We have investigated for the first time magnetocapacitance signatures modulated by DC bias applied in InAs/GaSb quantum wells, separated by a $20$\,nm AlSb barrier, realized by independent ohmic contacts with respect to InAs and GaSb channels. We probed an oscillatory behavior of capacitance as a function of magnetic field, representing the density of states of the confined two dimensional electron and hole charge carrier systems formed in the InAs and GaSb QW, respectively, in good agreement with our simulation. This oscillatory behavior shifts with DC bias applied, remarking the adjustable charge carrier density via bias voltage. In particular for forward bias greater than $0.02$\,V, we observed the formation of beating signatures, defining two different densities coexisting in the system, supported by the power spectra analysis performed for zero and non zero Zeeman spin splitting magnetic field range. The unique Landau phase diagram exhibits exceptional agreement with filling factor expectation curves either for even or odd parity, indicating strong intermixing effect due to crossing between two densities coexisting in the system, carrying spin states information as well. A huge Rashba coefficient and \textit{g}-factor also points out the relevance of spin-related properties predicted by theory.

The potential impact of capacitance measurements is expandable \cite{PhysRevB.60.R13958,capacitanceoscill_PhysRevB.108.075301,PhysRevLett.78.4613,gfactor_capacitance_Irie_2019}.
It can lead to the capability to access physical parameters and contribute to our understanding of excitons, applicable towards infrared laser detectors \cite{largeSOIPhysRevB.61.16743}. Understanding how the electric field influences the charge carriers in each channel is a significant aspect towards the accessibility of a stable excitonic condensate \cite{naveh_barrier,naveh1995band}. The magnetocapacitance experiment presented in this paper is a novel technique to approach such a research environment. The spin orbit interaction also carry attractive features creating a promising and relevant platform to explore spin orbit properties in the InAs/AlSb/GaSb systems \cite{Muraki_QSHI_10.1063_1.4967471}, enabling the investigation of further physical aspects towards spin-based devices as spin field effect transistors. The modulation of spin up and down states are even more interesting to speculate the formation of persistent spin helix in this bilayer system, so far only observed for GaAs quantum wells \cite{PSH_Koralek_2009}.


\section{Acknowledgement} 
The authors acknowledge financial support from the Swiss National Science Foundation (SNSF). We thank Prof. Dr. T. Ihn and Dr. Z. Lei for profitable physical insights and P. Märki for providing the proper setup to measure capacitance. We thank ScopeM Center for Microscopy at ETH Zürich and Dr. P. Zeng for the infrastructure and lamella preparation, together with Sjoerd Telkamp and Dr. F. K\v{r}í\v{z}ek for providing the STEM raw images.


\FloatBarrier 


\begin{thebibliography}{400}

\bibitem{PhysRevLett.115.036803}
Qu, F., Beukman, A. J. A., Nadj-Perge, S., Wimmer, M., Nguyen, B.M., Yi, W., Thorp, J., Sokolich, M., Kiselev, A. A., Manfra, M.J., Marcus, C.M. and Kouwenhoven, L.P.,
\textit{Electric and Magnetic Tuning Between the Trivial and Topological Phases in InAs/GaSb Double Quantum Wells},
Phys. Rev. Lett., 115, 036803, 2015.

\bibitem{PhysRevLett.78.4613}
Yang, M. J., Yang, C. H., Bennett, B. R. and Shanabrook, B. V.,
\textit{Evidence of a Hybridization Gap in ``Semimetallic'' InAs/GaSb Systems},
Phys. Rev. Lett., 78, 4613--4616, 1997.

\bibitem{zakharova_PhysRevB.64.235332}
Zakharova, A., Yen, S. T. and Chao, K. A.,
\textit{Hybridization of electron, light-hole, and heavy-hole states in InAs/GaSb quantum wells},
Phys. Rev. B, 64, 235332, 2001. 

\bibitem{du2017evidence}
Du, L., Li, X., Lou, W., Sullivan, G., Chang, K., Kono, J. and Du, Rui-Rui,
\textit{Evidence for a topological excitonic insulator in InAs/GaSb bilayers},
Nature communications, 8, 1971, 2017. 

\bibitem{edgestates_PhysRevB.97.045420}
Li, Chang-An and Zhang, Song-Bo and Shen, Shun-Qing,
\textit{Hidden edge Dirac point and robust quantum edge transport in InAs/GaSb quantum wells},
Phys. Rev. B, 97, 045420, 2018. 

\bibitem{helicaledege_PhysRevLett.114.096802}
Du, Lingjie and Knez, Ivan and Sullivan, Gerard and Du, Rui-Rui,
\textit{Robust Helical Edge Transport in Gated InAs/GaSb Bilayers},
Phys. Rev. Lett., 114, 096802, 2015. 

\bibitem{Edgechannel_PhysRevB.87.235311}
Suzuki, K. and Harada, Y. and Onomitsu, K. and Muraki, K.,
\textit{Edge channel transport in the InAs/GaSb topological insulating phase},
Phys. Rev. B, 87, 235311, 2013. 

\bibitem{nichele2016edge}
Nichele, Fabrizio and Suominen, Henri J and Kjaergaard, Morten and Marcus, Charles M and Sajadi, Ebrahim and Folk, Joshua A and Qu, Fanming and Beukman, Arjan JA and De Vries, Folkert K and Van Veen, Jasper and others,
\textit{Edge transport in the trivial phase of InAs/GaSb},
New Journal of Physics, 18, 083005, 2016. 

\bibitem{susanne_PhysRevB.96.075406}
Mueller, Susanne and Mittag, Christopher and Tschirky, Thomas and Charpentier, Christophe and Wegscheider, Werner and Ensslin, Klaus and Ihn, Thomas,
\textit{Edge transport in InAs and InAs/GaSb quantum wells},
Phys. Rev. B, 96, 075406, 2017. 

\bibitem{lapushkin2004self}
Lapushkin, I and Zakharova, A and Yen, Shun-Tung and Chao, Koung-An,
\textit{A self-consistent investigation of the semimetal--semiconductor transition in InAs/GaSb quantum wells under external electric fields},
Journal of Physics: Condensed Matter, 16, 4677, 2004. 

\bibitem{PhysRevB.94.045317}
Hu, Lun-Hui and Liu, Chao-Xing and Xu, Dong-Hui and Zhang, Fu-Chun and Zhou, Yi,
\textit{Electric control of inverted gap and hybridization gap in type-II InAs/GaSb quantum wells},
Phys. Rev. B, 94, 045317, 2016. 

\bibitem{padilla2018confinement}
Padilla, J.L. and Medina-Bailon, C. and Alper, C. and Gamiz, F. and Ionescu, A.M.,
\textit{Confinement-induced InAs/GaSb heterojunction electron--hole bilayer tunneling field-effect transistor},
Applied Physics Letters, 112, 182101, 2018. 

\bibitem{LL_PhysRevB.69.115319}
Zakharova, A. and Yen, S. T. and Chao, K. A.,
\textit{Landau level structures and semimetal-semiconductor transition in strained InAs/GaSb quantum wells},
Phys. Rev. B, 69, 115319, 2004. 

\bibitem{sipahi_PhysRevB.104.195307}
de Medeiros, M. H. L., Teixeira, R.L.R. C., Sipahi, G. M., Dias da Silva, L.G.G.V.,
\textit{Electric field induced edge-state oscillations in InAs/GaSb quantum wells},
Phys. Rev. B, 104, 195307, 2021. 

\bibitem{QSHIproposalPhysRevLett.96.106802}
Bernevig, B. A., Zhang, S.C.,
\textit{Quantum Spin Hall Effect},
Phys. Rev. Lett., 96, 106802, 2006. 

\bibitem{Interplay_QSHI_PhysRevB.106.235421}
Paul, T., Becerra, V. F., Hyart, T.,
\textit{Interplay of quantum spin Hall effect and spontaneous time-reversal symmetry breaking in electron-hole bilayers. II. Zero-field topological superconductivity},
Phys. Rev. B, 106, 235421, 2022. 

\bibitem{QSHI_HgTe_InAs/GaSb}
A. Mawrie,
\textit{Magnetotransport properties of the Quantum Spin Hall and Quantum Hall states in an inverted HgTe/CdTe and InAs/GaSb quantum wells},
J. Phys.: Condens. Matter, 34, 245301, 2022. 

\bibitem{PhysRevLett.100.236601}
Liu, C., Hughes, T.L., Qi, X.L., Wang, K. and Zhang, S.C.,
\textit{Quantum Spin Hall Effect in Inverted Type-II Semiconductors},
Phys. Rev. Lett., 100, 236601, 2008. 

\bibitem{Muraki_QSHI_10.1063_1.4967471}
Akiho, T. and Couëdo, F. and Irie, H. and Suzuki, K. and Onomitsu, K. and Muraki, K.,
\textit{Engineering quantum spin Hall insulators by strained-layer heterostructures},
Applied Physics Letters, 109, 192105, 2016. 

\bibitem{SHE_PhysRevLett.100.056602}
Yang, Wen and Chang, Kai and Zhang, Shou-Cheng,
\textit{Intrinsic Spin Hall Effect Induced by Quantum Phase Transition in HgCdTe Quantum Wells},
Phys. Rev. Lett., 100, 056602, 2008. 

\bibitem{du2013observation}
Du, Lingjie and Knez, Ivan and Sullivan, Gerard and Du, Rui-Rui,
\textit{Observation of Quantum Spin Hall States in InAs/GaSb Bilayers under Broken Time-Reversal Symmetry},
arXiv preprint arXiv:1306.1925, 2013. 

\bibitem{tunable_gates_10.1063/1.118187}
Drndic, M. and Grimshaw, M. P. and Cooper, L. J. and Ritchie, D. A. and Patel, N. K.,
\textit{Tunable electron-hole gases in gated InAs/GaSb/AlSb systems},
Applied Physics Letters, 70, 481-483, 1997. 

\bibitem{nguyen2015high}
Nguyen, Binh-Minh and Yi, Wei and Noah, Ramsey and Thorp, Jacob and Sokolich, Marko,
\textit{High mobility back-gated InAs/GaSb double quantum well grown on GaSb substrate},
Applied Physics Letters, 106, 032107, 2015. 

\bibitem{Irie_PhysRevMaterials.4.104201}
H. Irie, T. Akiho, F. Couedo, K. Suzuki, K. Onomitsu, K. Muraki,
\textit{Energy gap tuning and gate-controlled topological phase transition in $\mathrm{In}\mathrm{As}/{\mathrm{In}}_{x}{\mathrm{Ga}}_{1\ensuremath{-}x}\mathrm{Sb}$ composite quantum wells},
Phys. Rev. Mater., 4, 104201, 2020. 

\bibitem{naveh_barrier}
Naveh, Y and Laikhtman, B,
\textit{Excitonic Instability and Electric-Field-Induced Phase Transition Towards a Two-Dimensional Exciton Condensate},
Physical review letters, 77, 900, 1996. 

\bibitem{naveh1995band}
Naveh, Y and Laikhtman, B,
\textit{Band-structure tailoring by electric field in a weakly coupled electron-hole system},
Applied physics letters, 66, 1980--1982, 1995. 

\bibitem{ZHU1990595}
Xiaodong Zhu and J.J. Quinn and Godfrey Gumbs,
\textit{Excitonic insulator transition in a GaSb-AlSb-InAs quantum-well structure},
Solid State Communications, 75, 595-599, 1990. 

\bibitem{Eisenstein2004BoseEinsteinCO}
J. P. Eisenstein and A. H. Macdonald,
\textit{Bose–Einstein condensation of excitons in bilayer electron systems},
Nature, 5, 691-694, 2004. 

\bibitem{laikhtman_de1999band}
S. de-Leon, L.D. Shvartsman, B. Laikhtman,
\textit{Band structure of coupled InAs/GaSb quantum wells},
Physical Review B, 60, 1861, 1999. 

\bibitem{hirayama_PhysRevB.67.195319}
Suzuki, K. and Miyashita, S. and Hirayama, Y.,
\textit{Transport properties in asymmetric InAs/AlSb/GaSb electron--hole hybridized systems},
Phys. Rev. B, 67, 195319, 2003. 

\bibitem{nextnano4294186}
Birner, Stefan and Zibold, Tobias and Andlauer, Till and Kubis, Tillmann and Sabathil, Matthias and Trellakis, Alex and Vogl, Peter,
\textit{nextnano: General Purpose 3-D Simulations},
IEEE Transactions on Electron Devices, 54, 2137-2142, 2007. 

\bibitem{nextnano_article}
Trellakis, Alex and Zibold, Tobias and Andlauer, Till and Birner, Stefan and Smith, R. and Morschl, Richard and Vogl, Peter,
\textit{The 3D nanometer device project nextnano: Concepts, methods, results},
Journal of Computational Electronics, 5, 285-289, 2006. 

\bibitem{ihn2009semiconductor}
Ihn, Thomas,
\textit{Semiconductor Nanostructures: Quantum states and electronic transport},
OUP Oxford, 2009. 

\bibitem{phase_slips_Matija_PhysRevB.99.201402}
Karalic, M. and Mittag, C. and Mueller, S. and Tschirky, T. and Wegscheider, W. and Ensslin, K. and Ihn, T. and Glazman, L.,
\textit{Phase slips and parity jumps in quantum oscillations of inverted InAs/GaSb quantum wells},
Phys. Rev. B, 99, 201402(R), 2019. 

\bibitem{Rashba_PhysRevB.55.R1958}
G. Engels, J. Lange, T. Sch\"apers, H. L\"uth, 
\textit{Experimental and theoretical approach to spin splitting in modulation-doped ${\mathrm{In}}_{\mathrm{x}}$${\mathrm{Ga}}_{1\mathrm{\ensuremath{-}}\mathrm{x}}$As/InP quantum wells for B\ensuremath{\rightarrow}0},
Phys. Rev. B, 55, 1958(R), 1997. 

\bibitem{spinstates_PhysRevB.80.035303}
Li, Jun and Yang, Wen and Chang, Kai,
\textit{Spin states in InAs/AlSb/GaSb semiconductor quantum wells},
Phys. Rev. B, 80, 035303, 2009. 

\bibitem{RSS_10.1063/1.2909544}
Li, Jun and Chang, Kai and Hai, G. Q. and Chan, K. S.,
\textit{Anomalous Rashba spin-orbit interaction in InAs/GaSb quantum wells},
Applied Physics Letters, 92, 152107, 2008. 

\bibitem{gfactor_capacitance_Irie_2019}
Hiroshi Irie and Takafumi Akiho and Koji Muraki,
\textit{Determination of g-factor in InAs two-dimensional electron system by capacitance spectroscopy},
Applied Physics Express, 6, 063004, 2019. 

\bibitem{PhysRevB.60.R13958}
Yang, M. J. and Yang, C. H. and Bennett, B. R.,
\textit{Magnetocapacitance and far-infrared photoconductivity in GaSb/InAs composite quantum wells},
Phys. Rev. B, 60, R13958--R13961, 1999. 

\bibitem{capacitanceoscill_PhysRevB.108.075301}
Minkov, G. M. and Rut, O. E. and Sherstobitov, A. A. and Dvoretski, S. A. and Mikhailov, N. N. and Germanenko, A. V.,
\textit{Quantum oscillations of transport coefficients and capacitance: A manifestation of the spin Hall effect},
Phys. Rev. B, 108, 075301, 2023. 

\bibitem{largeSOIPhysRevB.61.16743}
Halvorsen, E. and Galperin, Y. and Chao, K. A.,
\textit{Optical transitions in broken gap heterostructures},
Phys. Rev. B, 61, 16743--16749, 2000. 

\bibitem{PSH_Koralek_2009}
Koralek, J. D. and Weber, C. P. and Orenstein, J. and Bernevig, B. A. and Zhang, Shou-Cheng and Mack, S. and Awschalom, D. D.,
\textit{Emergence of the persistent spin helix in semiconductor quantum wells},
Nature, 458, 610–613, 2009. 

\end{thebibliography}
\end{document}


\title{Supplemental Material: Magnetocapacitance oscillations dominated by giant Rashba spin orbit interaction in InAs/GaSb quantum wells separated by AlSb barrier}

\author{A.S.L. Ribeiro}
\author{R. Schott}
\author{C. Reichl}
\author{W. Dietsche} 
\author{W. Wegscheider} 
\affiliation{Advanced Semiconductor Quantum Materials Group, Eidgenössische Technische Hochschule ETH, Zürich, Switzerland}
\affiliation{Quantum Center, ETH Zürich, Switzerland}
\date{\today}

\maketitle

\section{Growth details}

The InAs/AlSb/GaSb heterostructure depicted in Fig. 1 from the main text, was grown using a modified Veeco Gen II molecular beam epitaxy (MBE) system, equipped with valved cracker cells for the group V component of the compound semiconductors. The protective oxide layer was desorbed from the undoped GaSb [$001$]-oriented substrate at temperatures of $540^\circ$C, under antimony counter pressure. To ensure precise temperature measurement and control during growth, a pyrometer system with black-body radiation fitting was employed. Growth rates were calibrated using reflection high-energy electron diffraction (RHEED) oscillations, with growth rates of $2$\,$\text{\AA}$/s for GaSb, $1$\,$\text{\AA}$/s for AlSb, and $0.4$\,$\text{\AA}$/s for InAs. Group V/III flux ratios of approximately $3.5$ for GaSb, $7$ for AlSb, and $5$ for InAs were applied.

After complete desorption of the oxide and clear visibility of typical GaSb RHEED patterns, growth was initiated with a buffer layer consisting of $50$\,nm GaSb, a $10$-period Al$_{0.33}$Ga$_{0.67}$Sb/GaSb superlattice, and $250$\,nm GaSb grown at $520^\circ$C (Fig. 1). For more effective p-type doping with silicon (Si) in the following $100$\,nm thick GaSb layer, the substrate temperature was increased to $535^\circ$C, achieving a doping concentration of $2\times 10^{19}$ cm$^{-3}$. Subsequently, the substrate temperature was lowered to $515^\circ$C, and $50$\,nm GaSb and $20$\,nm AlSb were grown. During the last $7$\,nm of the $20$\,nm thick AlSb barrier, the substrate temperature was lowered to $425^\circ$C to improve interface qualities and reduce defects. A $50$\,nm InAs layer was grown using a shutter sequence proposed by Tuttle et al. \cite{tutle}. Following the shutter sequence, the arsenic (As) pressure was slowly increased from a V/III ratio close to one to the final ratio of 5, in order to prevent the formation of void defects associated with As etching of the Sb-based layer underneath \cite{PhysRevMaterials}.

\section{I-V trace}

We display the electrical current as a function of DC bias voltage response measured from p to n layer in Fig. \ref{fig: IV trace}. 

\begin{figure}[htbp]
\centering
    \includegraphics[width = 0.8\columnwidth]{Figures/IV trace Odin.pdf}
    \caption{\textbf{I-V trace.} A central insulating low conductive region, of maximal resistance $\approx100$\,k$\Omega$.}
    \label{fig: IV trace}
\end{figure}

\section{\textit{g} - factor}

The determined \textit{g}-factor as a function of DC bias voltage is presented in Fig. \ref{fig:gfactor}. 

\begin{figure}[htbp]
\centering
    \includegraphics[width = 0.8\columnwidth]{Figures/final g factor with error bars.pdf}
    \caption{\textbf{\textit{g}-factor.} We depict the g-factor determined from the occurence of the first magnetocapacitance oscillations and Zeeman spin splitting for each DC bias voltage.}
    \label{fig:gfactor}
\end{figure}